\documentclass[preprintnumbers,showpacs,amsmath,amssymb,floatfix,prd,twocolumn,superscriptaddress,nofootinbib]{revtex4}
\usepackage{graphicx}
\usepackage{epsfig}
\usepackage{bm}
\usepackage{amsfonts}
\usepackage{epstopdf}
\usepackage{bm}

\begin{document}

\title{ Is Cosmological Constant Needed in Higgs Inflation?}

\author{Chao-Jun Feng}
\email{fengcj@shnu.edu.cn} 
\affiliation{Shanghai United Center for Astrophysics (SUCA), \\ Shanghai Normal University,
    100 Guilin Road, Shanghai 200234, P.R.China}
    \affiliation{State Key Laboratory of Theoretical Physics, \\Institute of Theoretical Physics, Chinese Academy of Sciences, Beijing 100190, P.R.China}

\author{Xin-Zhou Li}
\email{kychz@shnu.edu.cn} \affiliation{Shanghai United Center for Astrophysics (SUCA),  \\ Shanghai Normal University,
    100 Guilin Road, Shanghai 200234, P.R.China}

\begin{abstract}
The detection of B-mode shows a very powerful  constraint to theoretical inflation models through the measurement of the tensor-to-scalar ratio $r$. Higgs boson is the most likely candidate of the inflaton field. But usually, Higgs inflation models predict a small value of $r$, which is not quite consistent with the recent results from BICEP2. In this paper, we explored whether a cosmological constant energy component is needed to improve the situation. And we found the answer is yes. For the so-called Higgs chaotic inflation model with a quadratic potential, it predicts $r\approx 0.2$, $n_s\approx0.96$ with e-folds number $N\approx 56$, which is large enough to overcome the problems such as the horizon problem in the Big Bang cosmology. The required energy scale of the cosmological constant is roughly $\Lambda \sim (10^{14} \text{GeV})^2 $, which means a mechanism is still needed to solve the fine-tuning problem in the later time evolution of the universe, e.g. by introducing some dark energy component.
\end{abstract}

\maketitle


\section{Introduction}

Recently the detection of B-mode from CMB by the BICEP2 group \cite{Ade:2014xna} has indicated a strong evidence of inflation \cite{Guth:1980zm,Linde:1981mu,Albrecht:1982wi}, which solves many theoretical puzzles in the Big Bang cosmology.  The B-mode polarization can be only generated by the tensor perturbations. According to the reports of the BICEP2 experiment, the tensor-to-scalar ratio is in range: $r=0.20_{-0.05}^{+0.07} (68\%$ CL). 

In a simplest slow-roll inflation model, the early universe was driven by a single scalar field $\phi$ with a very flat potential $V(\phi)$. Usually, we call this field the inflaton. Although there are many inflation models in the market, we still do not well-understand what is the inflaton. The most economical and fundamental candidate for the inflaton is the standard model (SM) Higgs boson, which has been already observed by the collider experiment LHC in 2012 \cite{Aad:2012tfa,Chatrchyan:2012ufa}.  In this sense, Higgs inflation is a simple and elegant model. However, it is not easy for the Higgs boson to realize a inflation model with correct density perturbations.  To see this, we estimate the inflaton mass from the amplitude $A_s$ of the scalar perturbation power spectrum in the chaotic inflation model \cite{Linde:1983gd} with a quadratic  potential $V(\phi)=m^2\dot\phi^2/2$:
\begin{equation}\label{equ:mass1}
	m \approx 1.5\times 10^{13}\left(\frac{N}{60}\right)^{-1}\left( \frac{ 10^{9}A_s }{2.19}\right)^{1/2} \text{GeV} \,,
\end{equation}
which is many orders of magnitude larger than the observed Higgs mass, $m_{h}\approx 125.9\pm0.4$ GeV. In other words, the potential of Higgs field $h$ is not flat enough to realize an inflation. By introducing a non-minimal coupling to the gravity ($\sim h^2 R$) ,  one could indeed achieve such a flat potential \cite{Bezrukov:2007ep} after a conformal transformation.  And the predictions of this kind of non-minimal coupling Higgs inflation  are well consistent with observations before BICEP2. The authors in ref.\cite{Cook:2014dga} have found that this model can not accommodate the new measurement from BICEP2, because it generally predicts a small amplitude of tensor perturbations. An alternative Higgs inflation model was proposed in ref.\cite{Germani:2010gm}, in which the Higgs boson kinetic term is non-minimally coupled to the Einstein tensor ($\sim G^{ab}\partial_a h\partial_b h$).  According to the recent analysis on this model \cite{Germani:2014hqa}, it predicts $r\approx0.16$ when the number of e-folds $N\approx33$, since $r\approx16/(3N+1)$ in this model. However, to overcome the problems in the Big Bang theory, the number of e-folds is required to be around $N\approx60$, then the tensor-to-scalar ratio becomes even smaller, say $r\approx 0.09$. 

Another interesting Higgs inflation model called the Higgs chaotic inflation is proposed in ref.\cite{Nakayama:2010sk}. In this model, the SM Higgs boson realizes the quadratic chaotic inflation model, based on the so-called running kinetic inflation \cite{Nakayama:2010kt, Takahashi:2010ky}. The kinetic term of the inflaton is significantly modified at large field values, while it becomes the canonical one when $h$ is small. The  value of $r$ in this model is the same as that in the chaotic inflation model with a quadratic  potential, i.e. $r=8/N$. For $N\approx 60$, it predicts $r \approx 0.13$, but if we require a larger $r$, say $r\approx 0.2$, a smaller $N$ is needed, say $N \approx 40$, which is a little better than that predicted in the other Higgs inflation models, see ref.\cite{Nakayama:2014koa} for recent revisited in this model. It seems that the Higgs chaotic inflation is a charming Higgs inflation model in the market. 

On the other hand, there is a challenge for a single field inflation with BICEP2 result. For the chaotic inflation, the larger the value of the tensor-to-scalar ratio is, the smaller the value of the running of the spectral index is, see the details in ref.\cite{Gong:2014cqa}. Therefore, to be  more consistent  with observations, one might consider a little more beyond a single inflation model. Among many choices,  the cosmological constant is often forgotten when one building an inflation model, since by itself  only  the exact scale-invariant Harrizon-Zel'dovish power spectrum with the scalar spectral index $n_s =1$ could be produced, which is already ruled out at over $5\sigma$ by \textit{Planck} \cite{Ade:2013uln}. However, we find that the situation is changed when  the early universe is dominated by the cosmological constant as well as the inflaton.  It could give $n_s \approx 0.96$, $r\approx 0.2$ when the number of e-folds is not so small, say $N\approx 56$, and it also predict the correct magnitude of the spectrum amplitude. 

In the following, we will assume that the running kinetic approach is a correct way to realize inflation by SM Higgs boson and we also assume that both inflaton and the cosmological constant dominated the universe during the inflation time. In next section, we give a briefly review of the running kinetic inflation and then we pursue the role played by the cosmological constant during inflation. Finally, we will draw our conclusions and give some discussions in the last section.

\section{Running kinetic inflation}

The running kinetic inflation can be easily implemented in supergravity by assuming a shift symmetry exhibiting itself in the K\"ahler potential at high energy scales, while this symmetry is explicitly broken and therefore becomes much less prominent at low energy scales. In the unitary gauge, one can write down the Lagrangian for the Higgs boson $h$ \cite{Nakayama:2010sk, Nakayama:2010kt, Takahashi:2010ky, Nakayama:2014koa}:
\begin{equation}\label{equ:lag}
	\mathcal{L} = \frac{1}{2} \left(1+ \xi \frac{h^2}{2} \right) (\partial h)^2 - \frac{\lambda_h}{4} (h^2 - v^2)^2 \,.
\end{equation}
The effect of non-canonical kinetic term is significant for large $h\geq 1/\sqrt{\xi}$.  The kinetic term grows as $h^2$, that is why the name ``running kinetic inflation''.   By redefining the Higgs field, one can rewrite the Lagrangian in terms of canonically normalized field $\phi \equiv \sqrt{\xi/8} h^2$ with the effective potential 
\begin{equation}\label{equ:potential}
	V(\phi) = \frac{1}{2} m^2 \phi^2 \,, \quad m^2 \equiv \frac{4\lambda_h}{\xi} M_{pl}^2 \,.
\end{equation}
Thus, the quadratic chaotic inflation occurs.

\section{The role of the cosmological constant during inflation}

Assuming the universe was dominated by both the inflaton and the cosmological constant, the Friedmann equation could be written as
\begin{equation}\label{equ:fried}
	3M_{pl}^2 H^2 \approx \frac{1}{2}m^2\phi^2 + \Lambda M_{pl}^2 \,,
\end{equation}
where $M_{pl} = (8\pi G)^{-1/2} \approx 2.435\times 10^{18}$ GeV is the reduced Planck mass.  Then by using definition of the slow-roll parameters, we get
\begin{eqnarray}
	\epsilon &\equiv& -\frac{\dot H}{H^2} =\frac{2m^4M_{pl}^2\phi^2}{ ( m^2\phi^2 + 2\Lambda M_{pl}^2)^2} \,, \label{equ:sl1}\\
	\eta &\equiv& -\frac{\ddot\phi}{\dot\phi H}-\epsilon  = \frac{ 2m^2 M_{pl}^2}{m^2\phi^2 + 2\Lambda M_{pl}^2} \,. \label{equ:sl2} 
\end{eqnarray}
And also  the amplitude of the scalar perturbation power spectrum is given by
\begin{equation}\label{equ:amp}
	A_s \approx  \frac{m^2\phi^2 + 2\Lambda M_{pl}^2}{48\pi^2 M_{pl}^4 \epsilon} \,,
\end{equation}
which is defined as $\mathcal{P}_{s} = A_s (k/k_*)^{n_s-1 + \cdots }$. By using the relations $n_s-1 = 2\eta - 6\epsilon$ with $n_s$  the scalar spectrum index and  $r=16\epsilon$ with $r$ the tensor-to-scalar ratio,   we obtain the inflaton mass in terms of $n_s, r $ and $A_s$:
\begin{equation}\label{equ:fmass}
	\frac{m^2}{M_{pl}^2}  = \frac{3\pi^2}{4} \left( n_s-1 + \frac{3r}{8}\right) r A_s \,,
\end{equation}
and also the value of the cosmological constant:
\begin{equation}\label{equ:fcc}
	\frac{\Lambda}{M_{pl}^2} = \frac{3\pi^2}{2} rA_s \left[ 1 - \frac{r}{8(n_s-1) + 3r} \right] \,.
\end{equation}
The  number of e-folds could be also given by
\begin{equation}\label{equ:fn}
	N \equiv \int H dt
	\approx \frac{\phi^2}{4 M_{pl}^2} + \frac{\Lambda}{m^2} \ln \left( \frac{\phi}{M_{pl}}\right) \,.
\end{equation}
By using  Eqs.(\ref{equ:fcc}), (\ref{equ:fn}) and the value of $\phi$: 
\begin{equation}\label{equ:fphi}
	\phi = \sqrt { \frac{r}{2} } \left( n_s-1 + \frac{3r}{8}\right)^{-1} M_{pl} \,.
\end{equation}
obtained from Eqs.(\ref{equ:sl1}), (\ref{equ:sl2}) and (\ref{equ:amp}), we get
\begin{widetext}
\begin{equation}\label{equ:fn2}
	N \approx   \left(n_s-1+\frac{3r}{8}\right)^{-2} \bigg\{ \frac{r}{8}  
	+ \left(n_s-1+\frac{r}{4} \right) \left[ \ln r - \ln2 - 2\ln  \left(n_s-1+\frac{3r}{8}\right)\right] \bigg\} \,.
\end{equation}
\end{widetext}
 
Substituting the observed values of $n_s \approx 0.96$, $r\approx0.20$ and $A_s \approx 2.19\times 10^{-9}$ into Eq.(\ref{equ:fmass}), we estimated the mass of the inflaton as $m \approx 2.59 \times 10^{13} $ GeV.  If $\xi$ is sufficiently large, say $\xi \approx 4.6\times 10^9$ in Eq.(\ref{equ:potential}), the quartic coupling could be  $\lambda_h \approx 0.13$, which is required to explain the correct electroweak scale and the Higgs boson mass $m_h=\sqrt{2\lambda_h} v$. The large value of $\xi$ could be understood in terms of symmetry, see refs.\cite{Nakayama:2010sk, Nakayama:2010kt, Takahashi:2010ky, Nakayama:2014koa} for details. 

The scale of the cosmological constant can be estimated from Eq.(\ref{equ:fcc}), $\Lambda \approx 1.85\times 10^{-9} M_{pl}^2\approx (1.05\times 10^{14} \text{GeV})^2$. As usual, the fine-tuning problem of the cosmological constant  still exist at later time. Alternatively, one can consider some dynamical  dark energy models instead, which are more like a cosmological constant  component at early time.  

\begin{figure}[h]
\begin{center}
\includegraphics[width=0.45\textwidth,angle=0]{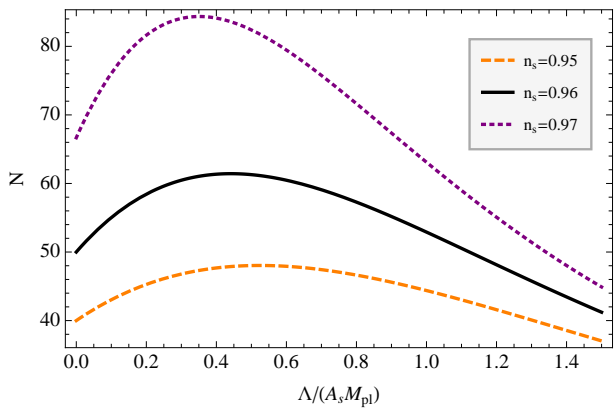}
\caption{\label{fig:nl}The number of e-folds $N$  v.s.  the cosmological constant $\Lambda$, which is measured in the unit of $A_sM_{pl}$. The dashed-orange, solid-black, and  dotted-purple curves correspond to $n_s = 0.95, 0.96, 0.97$ respectively.      }
\end{center}
\end{figure}

From Eq.(\ref{equ:fn2}), we obtain the number of e-folds as $N \approx 56$ , which looks enough to solve the horizon problem, the flat problem etc.  in the Big Bang cosmology. In other words, the model could predict $r\approx 0.2$ and $n_s\approx 0.96$ by given $N\approx 56$. Of course, the cosmological constant and the mass of the inflaton should take the values estimated above. From Fig.\ref{fig:nl}, one can see that  the value of $N$ increases with $\Lambda$  for small $\Lambda$ values, while it decreases for large $\Lambda$ values. This could be easy to understand: when $\Lambda$ is small, we have $\phi^2, m^2  \sim \Lambda$, see Eqs.(\ref{equ:fmass}), (\ref{equ:fcc}) and (\ref{equ:fphi}), then $N \sim \Lambda$.  But when $\Lambda$ is large, we have $\phi^2\sim 1/\Lambda, m^2\sim \Lambda^2$, then $N\sim 1/\Lambda$, which approaches to zero when $\Lambda$ goes to infinity.   

The latest analysis of the data including the $Planck$ CMB temperature data, the WMAP large scale polarization data  (WP) , CMB data extending the \textit{Planck} data to higher-$l$,  the \textit{Planck} lensing power spectrum, and BAO data  gives the constraint on the index $n_s$ of the scalar power spectrum\cite{Ade:2013uln}: $ 0.9583\pm0.0081$(\textit{Planck}+ WP), $ 0.9633\pm0.0072$ (\textit{Planck}+WP+lensing), $ 0.9570\pm0.0075$ (\textit{Planck}+WP+highL), $ 0.9607\pm0.0063$ (\textit{Planck} +WP+BAO).  It also gives an upper bound on $r\lesssim 0.25$.  The BICEP2 experiment constraints the tensor-scalar-ratio as: $r=0.20^{+0.07}_{-0.05}$ in ref.\cite{Ade:2014xna}. They are also other groups have reported their constrain results on the ratio: $r=0.23^{+0.05}_{-0.09}$ in ref.\cite{Cheng:2014ota} by adopting the Background Imaging of Cosmic Extragalactic Polarization (B2), \textit{Planck} and WP data sets;  $r=0.20^{+0.04}_{-0.05}$  in ref.\cite{Cheng:2014cja} combined with the Supernova Legacy Survey (SNLS); $r=0.199^{+0.037}_{-0.044}$ in ref.\cite{Li:2014cka}  by adopting the \textit{Planck}, supernova \textit{Union2.1} compilation, BAO and BICEP2 data sets; and also $r=0.20^{+0.04}_{-0.06}$ in ref.\cite{Wu:2014qxa} with other BAO data sets. This  B-mode signal can not be mimicked by topological defects\cite{Lizarraga:2014eaa}. The most likely origin of this signal is from the tensor perturbations or the  gravitational wave polarizations during inflation.

\begin{figure}[h]
\begin{center}
\includegraphics[width=0.45\textwidth,angle=0]{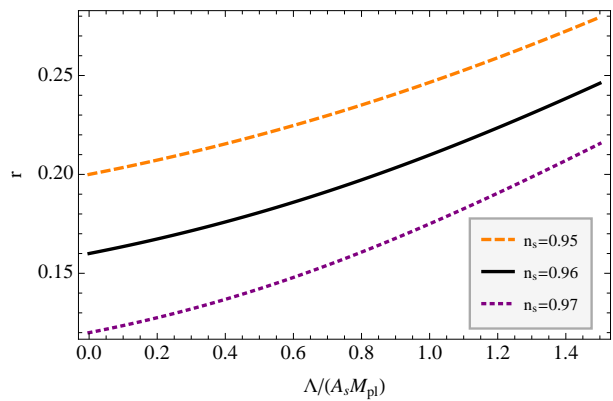}
\caption{\label{fig:rl}The tensor-to-scalar ratio $r$  v.s.  the cosmological constant $\Lambda$, which is measured in the unit of $A_sM_{pl}$. The dashed-orange, solid-black, and dotted-purple curves correspond to $n_s = 0.95, 0.96, 0.97$ respectively.   }
\end{center}
\end{figure}

Here, one can see that the cosmological constant plays an important role. It helps the universe to inflate at early time and contributes to the number of e-folds though Eq.(\ref{equ:fn}).  As a result, the inflaton field $\phi$ could be smaller than that without $\Lambda$.  To see this, we estimate $\phi \approx 9 M_{pl}$  from Eq.(\ref{equ:fphi}), while $\phi \approx \sqrt{4N} M_{pl} \approx 15 M_{pl}$ without $\Lambda$. Then, the slow-roll parameter $\epsilon$ could become also larger, which will then enhance the tensor-to-scalar ratio, $r\approx 16 \epsilon$, see Fig.\ref{fig:rl}. Therefore, it is likely that  the cosmological constant energy component is needed in the Higgs chaotic inflation with quadratic potential. 

\section{Conclusion and discussion}

The recent detection of B-mode by BICEP2 indicates an exciting leap forward in our ability to explore the early universe and fundamental physics. The measurement of the tensor-to-scalar ratio $r\approx0.2$ shows a very powerful constraint to theoretical inflation models. Higgs boson is the most likely candidate of the inflaton field. However, its mass $m_h\sim\mathcal{O}(10^2)$ GeV is much smaller  than that for a inflaton $m\sim\mathcal{O}(10^{13})$ GeV. To solve this  hierarchy  problem, a non-minimal coupling between the Higgs boson and gravity or a non-canonical kinetic term is needed. Usually, these Higgs inflation models predict a small value of $r$, which is not quite consistent with the results from BICEP2. In this paper, we explored whether a cosmological constant energy component is needed to improve the situation. And we found the answer is yes. The Higgs chaotic inflation now predicts $r\approx 0.2$, $n_s\approx0.96$ with e-folds number $N\approx 56$, which is large enough to overcome the problems in the Big Bang cosmology.  

However, we are still far from understanding the cosmological constant. And we haven't solve its fine-tuning problem in the later time evolution of the universe, which is asked why the present value of the cosmological constant is so small, or why the universe is accelerating at present $z\sim 1$. Noticed that the slow-roll parameters have a finite maximum value from Eqs.(\ref{equ:sl1}) and (\ref{equ:sl2}) as long as $\Lambda \neq 0$: $\epsilon_{\text{max}} \approx m^2/\Lambda$ when $\phi=\sqrt{2\Lambda}/m$, and $\eta_{\text{max}} \approx m^2/\Lambda$ when $\phi\rightarrow 0$. It seems that the inflation will never end if $\Lambda > m^2$.  To end the inflation, one may need a phase transition of a heavy Higgs boson $\chi$ with its mass at GUT scale, and it also  slightly couples to the light one  that responsible to inflation by $\sim h^2\chi^2$. At the beginning of inflation, the heavy Higgs boson is stable at its true vacuum ($\chi=0$), then it only contributes a constant potential, which can be regarded as the cosmological constant.  When the inflaton rolls down the potential and becomes small enough,   the vacuum  at $\chi=0$  turns to be a false one and the heavy boson would be no longer stable, then it rolls to  true vacuum to end the inflation.  In fact, the endless inflation is essentially due to the cosmological fine-tuning problem. Once a correct mechanism is found to  reduce $\Lambda$ to its present observational value, then the inflation would be certainly end. We will give a concrete example in detail to realize such a mechanism  that may solve the fine-tuning problem in the later work \cite{fengli}.

The challenge for a single field inflation to predict a large value of the running of the index still exit, $n_s' \equiv dn_s/d\ln k \approx -0.00025$ for $r\approx0.2$ in our case, see also ref.\cite{Gong:2014cqa} for detail discussions on this issue. But the constraint on the running is not so tight: $n_s'\approx-0.013\pm 0.009 (68\%\text{CL})$ from the analysis of \textit{Planck} data, see ref.\cite{Ade:2013uln}.  Furthermore, if  additional sterile neutrino species are taken into account in the universe, one could also obtain $r\approx0.20$ without the running of the spectral index ($n_s'\sim0$), see refs.\cite{Zhang:2014dxk,Dvorkin:2014lea,zhang1404.3598}. Certainly, if a large running is well-confirmed in future, then other mechanisms  explain it are urgently needed.

\acknowledgments

This work is supported by National Science Foundation of China grant Nos.~11105091 and~11047138, ``Chen Guang" project supported by Shanghai Municipal Education Commission and Shanghai Education Development Foundation Grant No. 12CG51, National Education Foundation of China grant  No.~2009312711004, Shanghai Natural Science Foundation, China grant No.~10ZR1422000, Key Project of Chinese Ministry of Education grant, No.~211059,  and  Shanghai Special Education Foundation, No.~ssd10004, and the Program of Shanghai Normal University (DXL124).


\begin{thebibliography}{999}


\bibitem{Ade:2014xna} 
  P.~A.~R.~Ade {\it et al.}  [BICEP2 Collaboration],
  arXiv:1403.3985 [astro-ph.CO].

\bibitem{Guth:1980zm} 
  A.~H.~Guth,
  Phys.\ Rev.\ D {\bf 23}, 347 (1981).

\bibitem{Linde:1981mu} 
  A.~D.~Linde,
  Phys.\ Lett.\ B {\bf 108}, 389 (1982).
  
\bibitem{Albrecht:1982wi} 
  A.~Albrecht and P.~J.~Steinhardt,
  Phys.\ Rev.\ Lett.\  {\bf 48}, 1220 (1982).

\bibitem{Aad:2012tfa} 
  G.~Aad {\it et al.}  [ATLAS Collaboration],
  Phys.\ Lett.\ B {\bf 716}, 1 (2012)
  [arXiv:1207.7214 [hep-ex]].
  
\bibitem{Chatrchyan:2012ufa} 
  S.~Chatrchyan {\it et al.}  [CMS Collaboration],
  Phys.\ Lett.\ B {\bf 716}, 30 (2012)
  [arXiv:1207.7235 [hep-ex]].
  
\bibitem{Linde:1983gd} 
  A.~D.~Linde,
  Phys.\ Lett.\ B {\bf 129}, 177 (1983).

\bibitem{Bezrukov:2007ep} 
  F.~L.~Bezrukov and M.~Shaposhnikov,
  Phys.\ Lett.\ B {\bf 659}, 703 (2008)
  [arXiv:0710.3755 [hep-th]].
  
  
\bibitem{Cook:2014dga} 
  J.~L.~Cook, L.~M.~Krauss, A.~J.~Long and S.~Sabharwal,
  arXiv:1403.4971 [astro-ph.CO].
  
\bibitem{Germani:2010gm} 
  C.~Germani and A.~Kehagias,
  Phys.\ Rev.\ Lett.\  {\bf 105}, 011302 (2010)
  [arXiv:1003.2635 [hep-ph]].
  
\bibitem{Germani:2014hqa} 
  C.~Germani, Y.~Watanabe and N.~Wintergerst,
  arXiv:1403.5766 [hep-ph].
  
\bibitem{Nakayama:2010sk} 
  K.~Nakayama and F.~Takahashi,
  JCAP {\bf 1102}, 010 (2011)
  [arXiv:1008.4457 [hep-ph]].

\bibitem{Nakayama:2010kt} 
  K.~Nakayama and F.~Takahashi,
  JCAP {\bf 1011}, 009 (2010)
  [arXiv:1008.2956 [hep-ph]].
  
  
\bibitem{Takahashi:2010ky} 
  F.~Takahashi,
  Phys.\ Lett.\ B {\bf 693}, 140 (2010)
  [arXiv:1006.2801 [hep-ph]].
    
     
  
  
\bibitem{Nakayama:2014koa} 
  K.~Nakayama and F.~Takahashi,
  arXiv:1403.4132 [hep-ph].
  
  
\bibitem{Gong:2014cqa} 
  Y.~Gong,
  arXiv:1403.5716 [gr-qc].
 
\bibitem{Ade:2013uln} 
  P.~A.~R.~Ade {\it et al.}  [Planck Collaboration],
  arXiv:1303.5082 [astro-ph.CO].
  
  
\bibitem{Cheng:2014ota} 
  C.~Cheng and Q.~G.~Huang,
  arXiv:1403.7173 [astro-ph.CO].
  
\bibitem{Cheng:2014cja} 
  C.~Cheng and Q.~G.~Huang,
  arXiv:1404.1230 [astro-ph.CO].
    
\bibitem{Li:2014cka} 
  H.~Li, J.~Q.~Xia and X.~Zhang,
  arXiv:1404.0238 [astro-ph.CO].
  
\bibitem{Wu:2014qxa} 
  F.~Wu, Y.~Li, Y.~Lu and X.~Chen,
  arXiv:1403.6462 [astro-ph.CO].
    
\bibitem{Lizarraga:2014eaa} 
  J.~Lizarraga, J.~Urrestilla, D.~Daverio, M.~Hindmarsh, M.~Kunz and A.~R.~Liddle,
  arXiv:1403.4924 [astro-ph.CO].
    
 \bibitem{fengli}
 C.~J.~Feng and X.~Z.~Li, work in progress.     
 
\bibitem{Zhang:2014dxk} 
  J.~F.~Zhang, Y.~H.~Li and X.~Zhang,
  arXiv:1403.7028 [astro-ph.CO].
  
\bibitem{Dvorkin:2014lea} 
  C.~Dvorkin, M.~Wyman, D.~H.~Rudd and W.~Hu,
  arXiv:1403.8049 [astro-ph.CO].
  
\bibitem{zhang1404.3598} 
  J.~F.~Zhang, Y.~H.~Li and X.~Zhang,
  arXiv:1404.3598 [astro-ph.CO].
 
 
\end{thebibliography}
\end{document}